\RequirePackage{fix-cm}
\documentclass[smallextended,twocolumn]{svjour3}       % onecolumn (second format)
\smartqed  % flush right qed marks, e.g. at end of proof
\usepackage{graphicx}
\usepackage{hyperref}
\usepackage{todonotes}
\begin{document}
% Prevent too long lines
\emergencystretch 3em

\title{A Conceptual Architecture for Educational Digital Twins Supporting AI Literacy Across Educational and Professional Settings
}

\titlerunning{Educational Digital
Twins Supporting AI Literacy }        % if too long for running head

\author{Antonio Bucchiarone \and Riccardo Belliato   \and Judith Michael \and Michael Mittermaier
}


\institute{Antonio Bucchiarone \at
              University of L'Aquila, Italy \\
              \email{antonio.bucchiarone@univaq.it}           %  \\
\and 
Riccardo Belliato \at
Fondazione Bruno Kessler (FBK), Trento, Italy \\
\email{rbelliato@fbk.eu}
           \and
           Judith Michael \at
              University of Regensburg, Germany\\
              \email{judith.michael@ur.de}
              \and
           Michael Mittermaier \at
              University of Regensburg, Germany\\
              \email{michael.mittermaier@ur.de}
}

\date{Received: date / Accepted: date}

\maketitle

\begin{abstract}
% Motivation
In the AI Literacy for Multidisciplinary Professional Readiness and Outreach (AIM-PRO) project, we are creating integrated methods to improve the education on AI literacy.
% Digital twins
One concept on which the project relies is educational digital twins, that is, digital representations of educator trainers, teachers, and learners that can be used in different stages of the educational process.
% impact
Such digital twins enable the simulation, monitoring, and optimization of learning experiences.
% what we present
This paper presents the AIM-PRO project and its conceptual foundations, focusing on its core objective: designing and implementing Digital Twins for Education to foster AI literacy across higher education, vocational education and training and professional learning environments.
%\subclass{MSC code1 \and MSC code2 \and more}
\end{abstract}

\section{Introduction}
In the era of digital transformation, societies are undergoing rapid and profound changes driven by advances in artificial intelligence (AI) technologies. These developments are reshaping economic sectors, professional practices, and everyday life, posing new challenges for education systems worldwide. As AI competency becomes increasingly vital for future societies \cite{Santana2023AICompetencies}, education must support the development of \emph{AI literacy}, an interdisciplinary competency that combines technical understanding with critical thinking and responsible use of intelligent technologies \cite{LongMagerko2020,Tondeur2025SQD}. Similarly to traditional literacy, such as reading, writing, and mathematical reasoning, AI literacy is emerging as a fundamental skill set that individuals must acquire to actively participate in the evolving digital society and remain competitive in the future workforce \cite{Kong2021AILiteracyCourse,Walter24,Biagini2025Towards}. At the same time, the rapid rise of generative artificial intelligence is opening new opportunities for education, enabling innovative pedagogical approaches and the design of novel learning environments that support adaptive and personalized learning experiences \cite{Murugesan2023GenAI,Yan2024Promises}.

AI literacy encompasses not only technical knowledge about algorithms
and data, but also an understanding of the ethical, societal, and environmental implications of AI technologies. Recent policy and religious initiatives, including the
European Union AI Act~\cite{ec_ai_regulatory_framework}, the Digital Education Action Plan~\cite{ec_digital_education_action_plan} and the encyclical letter Magnifica Humanitas~\cite{leoxiv_magnifica_humanitas}, emphasize
the importance of equipping citizens, professionals, and educators with the
competencies required to responsibly interact with AI systems and participate
in AI-driven innovation ecosystems.

Despite this growing importance, educational institutions and professional training providers face significant difficulties in integrating AI literacy into their curricula. Existing learning resources are often fragmented, inadequately adapted to different learning profiles, and disconnected from rapidly evolving technological developments. Furthermore, many educators lack adequate tools and pedagogical frameworks to design effective learning scenarios that integrate AI technologies in a meaningful and responsible way. These challenges are particularly evident in the diverse contexts of Higher Education (HE), Vocational Education and Training (VET), and professional training environments, where learners present heterogeneous backgrounds, skills, and learning needs. As a consequence, there is a growing need for integrated frameworks and technological infrastructures capable of supporting educators in the design, orchestration, and continuous improvement of AI-related learning experiences.

\begin{figure*}[htbp]
  \centering
  \includegraphics[width=\textwidth]{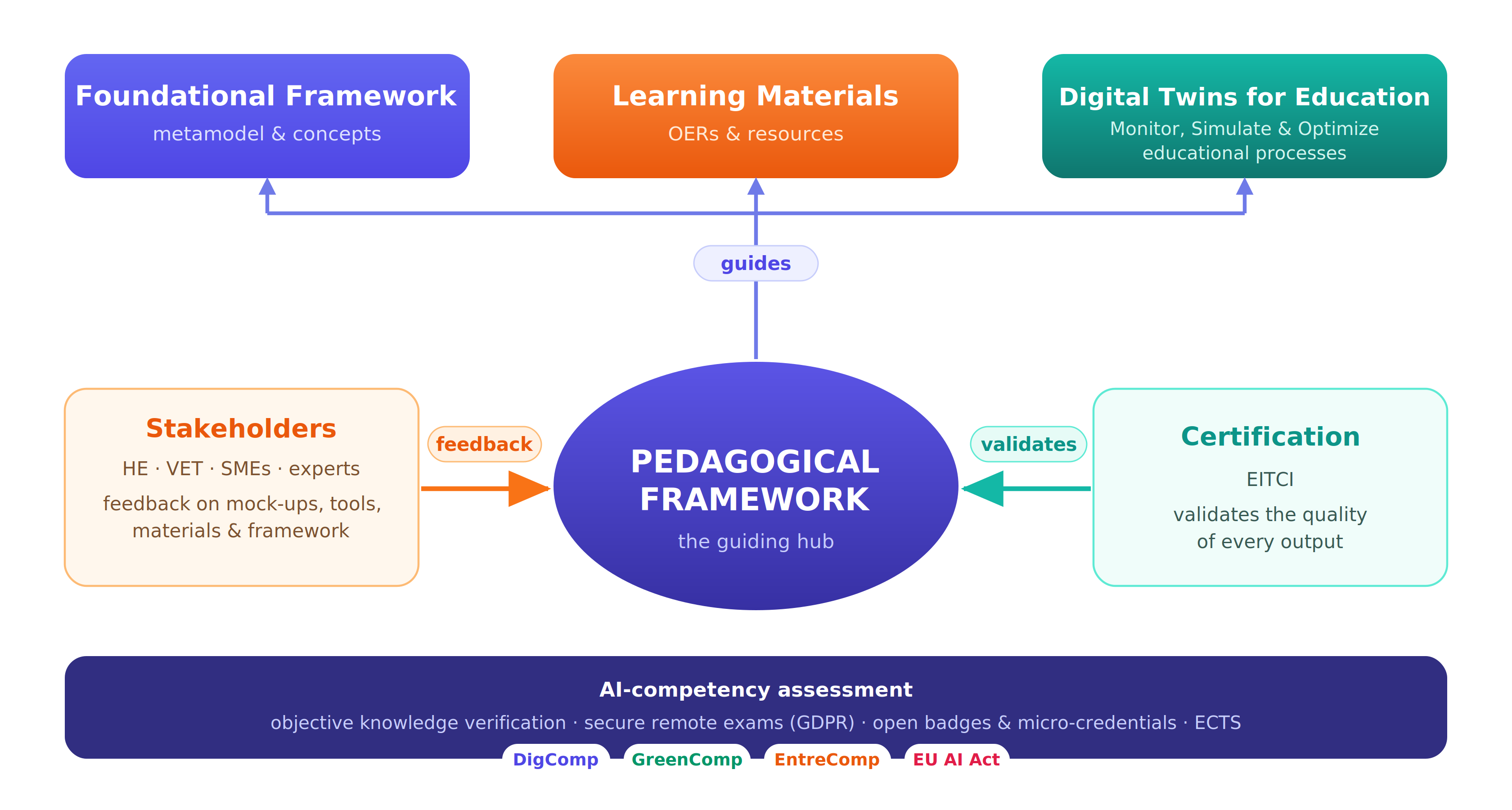}
  \caption{The AIM-PRO pedagogical framework: the guiding hub steers the project
  outputs, informed by stakeholder feedback and validated through certification.}
  \label{fig:aimpro-framework}
\end{figure*}

In response to these challenges, the \emph{AI Literacy for Multidisciplinary Professional Readiness and Outreach (AIM-PRO)} project~\cite{aimpro_project} proposes an integrated approach that combines pedagogical frameworks, open educational resources, generative AI technologies, and model-driven engineering techniques to support the development of AI literacy in educational and professional contexts. In particular, the project explores the use of \emph{digital twins for education} as a key enabling technology to monitor, simulate, and optimize 
learning environments and educational processes (as depicted in Figure \ref{fig:aimpro-framework}). By creating dynamic digital representations of teacher trainers, teachers, learners, learning activities, and educational contexts, Digital Twins can support data-driven pedagogical design and adaptive learning experiences.

The goal of this paper is therefore to present the AIM-PRO project and its conceptual foundations, with particular emphasis on its main objective of designing and implementing digital twins for education that support the development of AI literacy across higher education, vocational education and training, and professional learning environments.
%%%%%%%%%%%%%%%%%%%%%%%%%%%%%%%%%%%%%
\section{Overall project description }

AIM-PRO combines four complementary paradigms, open education technologies, generative AI, model-driven engineering (MDE), and digital twin engineering, to deliver adaptive, personalized learning experiences that build AI literacy across diverse contexts.

The scientific ambition of the project is to advance the conceptual and technological foundations of AI literacy education by addressing three main research objectives. First, the project aims to define a formalized \emph{foundational framework} for AI literacy learning scenarios. This framework combines pedagogical models with model-driven approaches to represent learning contexts, competencies, educational resources, and assessment mechanisms in a structured manner. Using modeling techniques, the project seeks to enable the systematic design, reuse, and adaptation of AI literacy learning scenarios in different educational settings.

Second, the project investigates the integration of advanced digital technologies into educational design and delivery. In particular, generative AI technologies are explored as tools for augmenting educational content and supporting educators in curating and adapting open educational resources (OERs). At the same time, digital infrastructures are developed to organize and structure educational resources related to AI literacy, sustainability, and entrepreneurship. These technologies enable the creation of dynamic and adaptive learning pathways that respond to the evolving needs and competencies of learners.

Third, the project explores the role of data-driven approaches in monitoring and improving educational processes. By combining learning analytics, adaptive feedback mechanisms, and digital simulation techniques, the project seeks to enable the continuous optimization of learning experiences. This approach allows educator trainers and teachers to experiment with pedagogical strategies, analyze learner engagement and progression, and iteratively refine educational practices.

Beyond these technical objectives, the project also emphasizes the importance of integrating ethical, environmental, and entrepreneurial perspectives into AI education. 

\textbf{How to reach the project objectives and main assets.} Currently, the project team is defining the AIM-PRO Foundational Framework, which establishes the conceptual models and technical requirements for AI literacy learning scenarios. 
Building upon these foundations, subsequent phases focus on the creation of educational resources repository and the development of digital tools to support learning design and delivery. These assets include an open database of AI literacy resources, generative AI tools for content creation, and modeling environments that allow educators to design structured learning scenarios.

A key technological asset of the project is the development of digital infrastructures capable of integrating these different components into a coherent ecosystem. These infrastructures allow educators to access curated resources, design personalized learning experiences, and monitor learning outcomes in multiple educational contexts. Certification mechanisms, including digital credentials and open badges, further support the recognition and validation of acquired AI competencies.

Finally, the tools and methodologies developed are validated through real-world implementations across higher education institutions, vocational education and training providers, and industrial partners. These pilot implementations enable iterative experimentation, feedback collection, and continuous refinement of the proposed approaches, ensuring both scientific validity and practical applicability.

\textbf{Transition to digital twin–based educational environments.}
Although the project integrates several complementary technological approaches, a central element of the AIM-PRO vision is the use of \emph{digital twins for education}. Digital twins (DTs)~\cite{tao2019five,kritzinger2018digital,PZC+25,BBM+24} enable the creation of dynamic virtual representations of learning environments, educator trainers, teachers, learners, and educational processes, allowing simulation, monitoring, and optimization of learning experiences. 

%The next section therefore focuses specifically on the role of Digital Twin technologies in education, explaining how they support adaptive learning, data-driven pedagogical design, and continuous improvement of AI literacy training environments.

\section{How DTs for education could help achieve the project goals}

\begin{figure}[htbp]
  \centering
  \includegraphics[width=\linewidth]{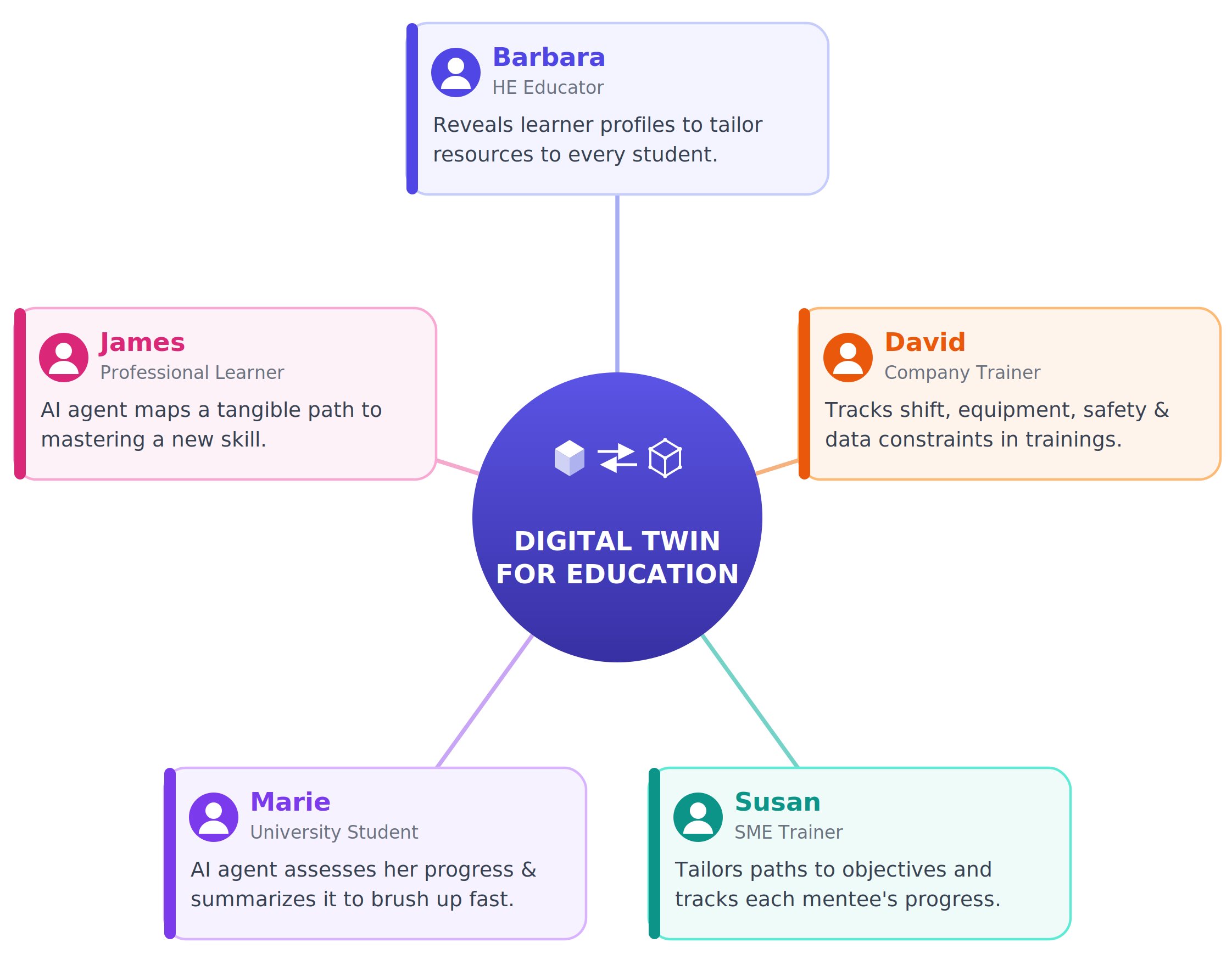}
  \caption{Digital twins supporting AI-literacy training across five stakeholder
  perspectives.}
  \label{fig:dt-scenarios}
\end{figure}

In AIM-PRO, our aim is to support the AI literacy training process (adaptive learning, data-driven pedagogical design, and continuous improvement of AI literacy training environments) for different stakeholders  with digital twins (as depicted in Figure \ref{fig:dt-scenarios}). Below, we sketch scenarios to show the pain points and solutions from different perspectives.

\textbf{An HE educator perspective.}
Barbara is a professor who plans a new class for the next semester. She is unclear on the level of skill of the students at the beginning of her class and on how to deliver the subjects to support different types of learning. The digital twin helps her understand the contexts and learning paths of the students in her class and helps her plan learning resources to make the content of her class accessible to different types of learners and different skill levels.

\textbf{A company trainer perspective.}
David supervises the continuous employee training in his company. He needs to ensure that organisational constraints (shift schedule, equipment availability) are met as well as company policies (safety rules, data protection) are enforced. The digital twin helps in keeping track of these constraints while planning and implementing long-term company trainings.

\textbf{A SME trainer perspective.}
Susan is a domain expert who trains individual junior team members in a small or medium-sized enterprise (SME) in short up-skilling sessions. With the digital twin, she can tailor the learning paths of her mentees to specific learning objectives and her preferences while tracking the individual learning progress of the team members.

\textbf{A student perspective at a university.}
Marie starts a postgraduate university degree. In her advanced software engineering class, there are  concepts that she has already learned in a software engineering class during her undergraduate studies. In the digital twin, an AI agent assesses her learning progress and presents a summary according to her learning type to efficiently brush up her skills before moving on to new concepts instead of spending more time than necessary on concepts she already knows.  

\textbf{A professional learner perspective.}
James is a professional learner with a couple of years of work experience. For a new project, he needs to master a new skill. He is uncertain of the best learning approach and overwhelmed by the variety of classes and learning resources. Within the digital twin, he evaluates his current abilities and his learning type, and the AI agent presents a tangible learning path linked to resources according to his learning type. This path of learning leads him from his current level to mastering the new skill. 

Clearly, these scenarios showcase only some of the possible perspectives. However, they already show that different stakeholders require a wide variety of services to support their educational processes, needs and personal preferences.

\section{How a Digital Twin Architecture for Education could look like}

\begin{figure*}
    \centering
    \includegraphics[width=1\textwidth]{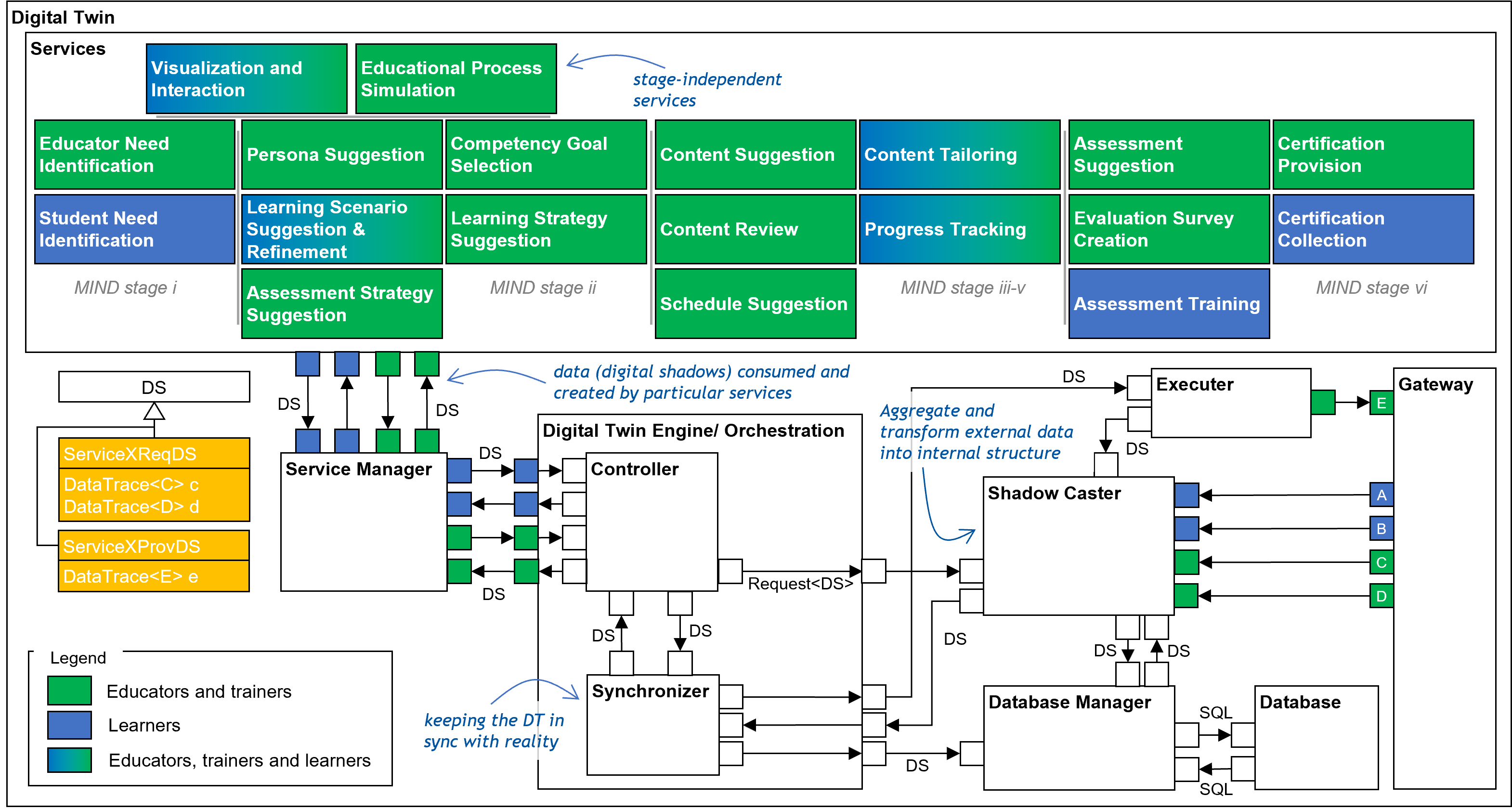}
    \caption{Conceptual architecture for a digital twin for education. Services in green are for educators, those in blue for learners, both colors are used where services would help both groups.}
    \label{fig:dt-architecture}
\end{figure*}

To realize DTs for education, we propose a conceptual architecture (see \autoref{fig:dt-architecture}) that can result in different variants of digital twins based on the selected services and implementation technologies. 
This architecture should support not only educator trainers, teachers, and learners, but also HE institutions and SMEs. 
This conceptual architecture does not place technical restrictions on the realization of a concrete educational DT: Monolithic, microservice-based, or a mix between agentic AI services and deductive services is possible. This is especially important because different institutions might already have tools in place for certain services that must be integrated in such an approach.
%Fig. \ref{fig:dt-architecture} provides an overview of this architecture.

The lower part of the DT structure in~\autoref{fig:dt-architecture} is slightly adapted from the conceptual DT architecture for manufacturing in~\cite{HMR+25}, that is based on the unifying reference model for digital twins of cyber-physical systems ~\cite{PZC+25}. 
The DT receives data from the actual system through the \texttt{Gateway}, which is processed by the \texttt{Shadow Caster} before being stored or sent further via the \texttt{DT Engine} or \texttt{Orchestrator} to \texttt{Services}. 
The DT engine includes a \texttt{Synchronizer} (to synchronize data from the actual system) and a \texttt{Controller} (for service and task orchestration). 
Via the \texttt{Executer}, the DT can send commands to the so-called actual system (in our case, educators and learners). In the context of education, this may include suggestions for the educator based on the current skill levels of the learners or direct updates and interaction with external learning tools via their APIs that then indirectly influence the educators or learners.

However, the main adaptions for the educational application domain are in the provided services sketched in the \texttt{Service Manager} that follow the six stages of the MIND model (see~\cite{monib2025mind}), an instructional design framework for creating AI-supported micro-learning experiences (in~\autoref{fig:dt-architecture} from left to right). In the MIND model, we start with (i) an analysis stage to identify needs (educators' and learners' needs), followed by (ii) a design stage where we define learning outcomes, identify potential personas, select essential content, assessment strategies, and mediality to design content.
In (iii) a development stage, we create and review content that is periodically delivered in (iv) a delivery stage. In (v) a receiving stage, the learners receive and practice content through interaction with peers and educators before, ultimately, we assess the learning outcomes, provide certificates, and evaluate the training (its content, mediality and how it helped improve their own competencies) in (vi) an assessment and evaluation stage.
Clearly, all these services need fitting visualizations and interaction formats, e.g., conversational interfaces for need identification from educators and learners or bar charts to show the learning progress for different skills. Another service that can be provided in several stages is a simulation of the educational process; it can be used in early stages with personas and later on based on the current progress of real learners. Such a simulation would enable educators to see whether the selected learning activities and strategies will lead to the desired results in improving certain skills.

However, this architecture is not restricted to the MIND model. If other pedagogical models shall be used, the modular combination of services allows for different orders and stage combinations, and new services can be easily added and registered in the service manager. If the DTs shall follow a particular educational process, it is reflected in the \texttt{DT Engine/Orchestrator} as order in which services can be called.

%personal skills, needs preferences
%organizational constraints

The proposed digital twin assists teachers and trainers in individually tailoring learning paths to learners according to their needs and skills while simultaneously following organizational constraints.

% What should we expand on? I think the services in the figure are named desdcriptive enough, but we could also list them in text and introduce them individually?
    % von links nach rechts nach dem mind model sortiert

% How much time should we spend on explaining the DT architecture around it?
    % Heithoff et al. 2025

%%%%%%%%%%
%main architectural decisions to take:
%either one DT or several DTs (different DT types)
%if several digital twins, we need connections/middleware between them

%concrete examples for HE/SMEs (2 figures?)

%a teacher that is new 

\section{What comes next}
As the AIM-PRO project started in February 2026, the main parts of the concrete realization of DTs for education are in development.
With our expertise in model-driven engineering~\cite{BGK+24,Bucchiarone20} and model-driven engineering of digital twins~\cite{KMR+20,FHM+23,HHMR23}, the core part of a digital twin for educators and learners can be developed from domain models that describe information needs in the different stages. Currently, these models are being developed together with experts in pedagogy and certification. With them as a base, the DT core and visualizations can be generated. In addition, there are currently different services under development, e.g., to capture educators needs, to tailor course material to different stakeholder needs, or to generate personalized learning paths.
Initial results are promising: the modular nature of these services suggests that the same DT core can be reused across stakeholder needs, reducing development effort while increasing pedagogical flexibility.
Taken together, these developments mark a first concrete step toward scalable, personalized AI literacy education and lay the groundwork for evaluating the architecture in real educational settings.

\section*{Funding}
This work was partially supported by the Erasmus+ Alliance for Innovation project AIM-PRO – AI Literacy for Multidisciplinary Professional Readiness and Outreach (Grant Agreement No. 101245864) and by Movetia, the Swiss agency promoting exchange, mobility and cooperation in education, continuing education, and youth work.

\section*{Conflict of Interest}
Judith Michael served as Guest Editor for this special issue. To avoid any conflict of interest, the editorial handling and peer review of this manuscript were managed independently by another guest editor. The authors declare no other competing interests.

%%%%%%%%%%%%%%%%%%%%%%%%%%%%%%%%%%%%%


\bibliographystyle{spmpsci}      % mathematics and physical sciences
\bibliography{bib/bib, bib/jm, bib/mm}   % name your BibTeX data base

%\begin{thebibliography}{}
% etc
%\end{thebibliography}

\end{document}